\documentclass[12pt, fleqn]{article}
\usepackage{latexsym}
\usepackage{amsmath}
\usepackage{amsfonts}
\usepackage{amssymb}
\usepackage{graphicx}
\usepackage{slashbox}
\usepackage{color}
\usepackage{tikz}
\usepgflibrary{arrows}

\title{Chiral fermions, \\massless particles and\\ Poincare covariance}
\author{
Krzysztof Andrzejewski\\Agnieszka Kijanka-Dec\\Piotr Kosi\'nski\thanks {e-mail: pkosinsk@uni.lodz.pl}\\ Pawe\l{} Ma\'slanka\\
\small \textit{Department of Computer Science} \\
\small \textit {Faculty of Physics and Applied Informatics}\\
\small \textit {University of Lodz}\\
\small \textit {Pomorska 149/153, 90-236 Lodz, Poland}
}

\date{}
\begin{document}
\maketitle
\begin{abstract}
The coadjoint orbit method is applied to the construction of Hamiltonian dynamics of massless particles of arbitrary helicity. The unusual transformation properties of canonical variables are interpreted in terms of nonlinear realizations of Poincare group. The action principle is formulated in terms  of new space-time variables with standard transformation properties. 
\end{abstract}
\section {Introduction} 
\par Recently, triangle anomalies, chiral fermions and Berry curvature in momentum space, their interrelations and role played in various physical phenomena have attracted much attention \cite{b1}-\cite{b23}. Much of the research consists in exploring anomaly-related phenomena in kinetic theory. An important point here is that, assuming weak external fields and weak particle interactions, one can rely to large extent on (semi)classical approximation. For example, instead of using the Weyl equation one can describe massless chiral fermions of helicity $\frac {1}{2}$ by the action functional
\begin {align} 
\label {al1}
S=\int \biggl ( \bigl(\vec {p}+e\vec {A}\bigr)\cdot \dot {\vec {x}}- \bigl(\mid \vec {p}\mid +e\Phi  \bigr )-\vec {\alpha }\cdot \dot {\vec {p}}\biggr ) dt
\end {align}
involving the vector potential $\vec {\alpha }(\vec {p})$ describing the Berry monopole in momentum space. Eq. (\ref{al1}) can be derived from Weyl Hamiltonian by considering semiclassical approximation to the path-integral representation of a transition amplitude \cite{b16} or, alternatively, using wave-packet approach \cite{b24}.
\par The main problem with eq.(\ref{al1}) is that it lacks manifest Lorentz symmetry even in the absence of external fields. It is the more surprising that it has been derived from explicitly covariant Weyl theory. To shed some light on the problem the authors of Ref. \cite{b16} proposed a modified transformation law for particle dynamical variables which is consistent in the sense that it leaves the dynamics following from the action (\ref{al1}) invariant and reduces to the standard Lorentz symmetry if the additional terms which arise due to the nonzero helicity are neglected. However, their proposal is exotic in the sense that: (i) it contains additional, helicity-dependent, terms mentioned above;\linebreak(ii) the group composition rule closes only \textquotedblleft on-shell\textquotedblright.
\par A deep analysis of the resulting situation has been given in the nice recent papers \cite{b14}, \cite{b22} and \cite{b23}. In particular, Duval et al. not only extended the results of Ref. \cite{b16} to the case of full Poincare symmetry but they reconsidered the whole problem in more general framework provided by the Souriau sympletic approach to dynamics \cite{b25}. They were able to derive the Poincare symmetry for chiral fermions by showing that the latter can be obtained from Souriau's model of relativistic massless spinning particle by the procedure called \textquotedblleft spin enslaving\textquotedblright.
\par Let us note that some of the apparently paradoxical features of Lorentz transformation laws for particles with nonzero spin (massive case) or helicity (massless case) appear to be unavoidable consequences of the group structure and basic conservation laws. It has been noticed long time ago \cite{b26} that the generators of Poincare symmetry for massless particle of nonzero helicity cannot be constructed out of canonical variables obeying standard canonical commutation rules and having standard transformation properties; if it were possible, the helicity would acquire more than one value within irreducible representation of Poincare group. Another nice argument in favour of \textquotedblleft exotic\textquotedblright transformation has been given in Ref. \cite{b16} (see also \cite{b15}) where the zero impact parameter collision of two massless particles of nonvanishing helicities was considered. By applying the Lorentz boost along the direction of motion of one incoming particle it is shown there that such a boost must result in \textquotedblleft side jump\textquotedblright \, in order to fulfil the angular momentum conservation law. Similar side jumps which depend only on the kinematics of the problem appear, for example, in impurity scattering caused by spin-orbit interaction \cite{b27}. This phenomenon seems also to have its counterpart in optics in the form of the relativistic Hall effect of light \cite{b28}-\cite{b34}, \cite{b15} (see also \cite{b35}-\cite{b37}).
\par The reason for the existence of the above described specific side jumps can be also traced back to the question of defining the center of mass of relativistic extended spinning body \cite{b38}, \cite{b39}, \cite{b14}.
\par In the present paper, inspired by the  Ref. \cite{b22}, we study further the invariance properties of the action functional (1). Our starting point is the construction of the Hamiltonian dynamics for massless particles with arbitrary helicity. The main tool we use is the coadjoint orbits method \cite{b25}, \cite{b40}-\cite{b42}; it has been already applied to the dynamics of relativistic particles in a number of references \cite{b25}, \cite{b43}-\cite{b45}, \cite{b46}-\cite{b49}. We classify the orbits corresponding to massless particles of the given helicities and construct the generators of Poincare group in terms of canonical variables. We find explicit form of the stability subgroup of a \textquotedblleft canonical\textquotedblright point on the orbit and reinterprete the whole construction in terms of nonlinear realizations of Poincare group. This allows for quite natural interpretation of \textquotedblleft exotic\textquotedblright transformation properties of coordinate variables. It is shown that the action principle can be put in the form which does not depend on the value of helicity; the latter enters only the transformation properties of basic variables. On the other hand, if one insists on having standard transformation properties of basic space-time variables, the action functional becomes helicity-dependent and exhibits the gauge symmetry, the gauge group being the stability subgroup mentioned above. The initial description is then obtained by an appropriate gauge fixing which is not covariant under the action of full Poincare group. Therefore, the action of the latter on initial variables is a composition of left action by standard space-time transformations supplemented by a gauge transformation. This provides alternative way of looking at the \textquotedblleft unusual\textquotedblright transformation properties of dynamical variables representing massless particles. 
\section{Classical massless particles}
\par We adopt the convention $g_{\mu \nu }=diag (+---)$. The light-cone coordinates are defined by $x^\pm =\frac{1}{\sqrt 2} (x^0\pm x^3)$. Let $k$ be fixed but arbitrary parameter having the momentum dimension and let $k^\mu =(k,0,0,k)$ be the standard null vector. Denote by $L_k\subset SO(3,1)$ the stability subgroup of $k^\mu $. Any element $\Lambda $ of $SO(3,1)$ can be decomposed as follows
\begin {align} 
\label {al2}
\Lambda =B\cdot D\cdot R, \quad \quad D\in L_k, \quad \quad R\in L_k
\end {align}
where in the light-cone basis $(x^+,x^-,x^1,x^2)$ the matrices $B$, $D$ and $R$ take the form
\begin {align} 
\label {al3}
B=\begin {pmatrix} \Lambda ^+_{\phantom{+}+}&0&0&0\\\Lambda ^-_{\phantom{-}+}&\frac{1}{\Lambda ^+_{\phantom{+}+}}&\frac {\Lambda ^1_{\phantom{1}+}} {\Lambda ^+_{\phantom{+}+}}&\frac {\Lambda ^2_{\phantom{2}+}} {\Lambda ^+_{\phantom{+}+}} \\ \Lambda ^1_{\phantom{1}+}&0&1&0\\\Lambda ^2 _{\phantom{2}+}&0&0&1\end {pmatrix}
\end {align}
\begin {align} 
\label {al4}
D=\begin {pmatrix} 1& \frac {\Lambda ^+_{\phantom {+}-}}{\Lambda ^+_{\phantom {+}+}}&d_1&d_2\\0&1&0&0\\0&d_1&1&0\\0&d_2&0&1\end {pmatrix}
\end {align}
\begin {align} 
\label {al5}
R=\begin {pmatrix}1&0&0&0\\0&1&0&0\\0&0&\cos \alpha &\sin \alpha \\0&0&-\sin\alpha &\cos\alpha  \end {pmatrix}
\end {align}
and 
\begin {align} 
\label {al6}
d_{1,2}\equiv \frac{\Lambda ^{1,2}_{\phantom{1,2}-}\Lambda ^+_{\phantom {+}+}-\Lambda^{1,2}_{\phantom{1,2}+}\Lambda^+_{\phantom {+}-}}{\Lambda^+_{\phantom {+}+}}
\end {align}
\begin {align} 
\label {al7}
\cos \alpha \equiv \frac{\Lambda ^{1}_{\phantom{1}1}\Lambda ^+_{\phantom {+}+}-\Lambda^{1}_{\phantom{1}+}\Lambda^+_{\phantom {+}1}}{\Lambda^+_{\phantom {+}+}}
\end {align}
\begin {align} 
\label {al8}
\sin \alpha \equiv \frac{\Lambda ^{1}_{\phantom{1}2}\Lambda ^+_{\phantom {+}+}-\Lambda^{1}_{\phantom{1}+}\Lambda^+_{\phantom {+}2}}{\Lambda^+_{\phantom {+}+}}
\end {align}
The decomposition (\ref{al2}) is singular at some points because the principal bundle $(SO(3,1),L_k)$ is nontrivial but this fact does not affect the reasoning. Note that $B$ parametrize the coset manifold $SO(3,1)/L_k$. Denote by $(\Lambda,a)$ the elements of Poincare group $\mathcal {P}$, the composition law being $(\Lambda,a)\cdot (\Lambda', a')=(\Lambda\Lambda', \Lambda a'+a)$. An infinitesimal element $g=(I+\omega ,\epsilon)$ can be written as 
\begin {align} 
\label {al9}
g=I+i\epsilon ^\mu P_\mu -\frac{i}{2}\omega ^{\mu \nu }M_{\mu \nu }
\end {align}
with $P_\mu $ and $M_{\mu \nu }=-M_{\nu \mu }$ being the generators for translations and Lorentz transformations, respectively. Denote by $\zeta _\mu $ and $\zeta _{\mu \nu }=-\zeta _{\nu \mu }$ the coordinates in the dual space to Lie algebra of $\mathcal {P}$. The coadjoint action of $\mathcal {P}$ reads
\begin {align} 
\label {al10}
Ad^*_{(\Lambda,a)}\zeta _\mu =\Lambda_\mu ^{\phantom {\mu }\nu }\zeta _\nu 
\end {align}
\begin {align} 
\label {al11}
Ad^*_{(\Lambda,a)}\zeta _{\mu \nu }=\Lambda_\mu ^{\phantom {\mu }\alpha  }\Lambda_\nu ^{\phantom {\nu  }\beta}\zeta _{\alpha \beta} +(a_\mu \Lambda_\nu ^{\phantom {\nu}\alpha }-a_\nu \Lambda_\mu ^{\phantom {\mu }\alpha })\zeta _\alpha 
\end {align}
The dual space is equipped with invariant Poisson structure which can be read off from the basic commutation rules of Poincare algebra:
\begin {align} 
\label {al12}
\{\zeta _\mu ,\zeta _\nu \}=0
\end {align}
\begin {align} 
\label {al13}
\{\zeta _{\mu \nu } ,\zeta _\alpha  \}=g_{\nu \alpha }\zeta _\mu -g_{\mu \alpha }\zeta _\nu 
\end {align}
\begin {align} 
\label {al14}
\{\zeta _{\mu \nu } ,\zeta _{\alpha \beta } \}= g_{\mu \beta }\zeta _{\nu \alpha }+g_{\nu \alpha }\zeta _{\mu \beta }-g_{\mu \alpha }\zeta _{\nu \beta }-g_{\nu \beta }\zeta _{\mu \alpha }
\end {align}
The coadjoint orbits are classified by selecting the values of the invariants corresponding to the Casimir operators
\begin {align} 
\label {al15}
\mathcal {M}^2\equiv \zeta ^\mu \zeta _\mu 
\end {align}
\begin {align} 
\label {al16}
\mathcal {W}^2\equiv w^\mu w _\mu\,, \quad \quad w ^\mu =\frac{1}{2}\epsilon ^{\mu \nu \alpha \beta }\zeta _\nu \zeta _{\alpha \beta}
\end {align}
Note the following Poisson brackets following from eqs. (\ref{al12}) - (\ref{al16}):
\begin {align} 
\label {al17}
\{w ^\mu ,\zeta _{\rho \sigma  }\}=\delta ^\mu _\rho w _\sigma  -\delta ^\mu _\sigma  w _\rho 
\end {align}
\begin {align} 
\label {al18}
\{w ^\mu ,w ^\nu \}=\epsilon ^{\mu \nu \rho \sigma  }\zeta _\rho w _\sigma   
\end {align}
We are interested in coadjoint orbits corresponding to $\mathcal {M}^2=0$, $\mathcal{W}^2=0$ and $\zeta ^0>0$. Due to the former condition any such orbit contains a point $(\underline{\zeta }\,_\mu, \,\underline {\zeta }\,_{\mu \nu })$ with $\underline {\zeta }\,_\mu =(k,0,0,-k)\equiv k_\mu$. Once $\underline {\zeta }\,_\mu $ is fixed, $\mathcal {W}^2=0$ yields
\begin {align} 
\label {al19}
\underline {\zeta }\,_{01}-\underline {\zeta }\,_{31}=0
\end {align}
\begin {align} 
\label {al20}
\underline {\zeta }\,_{02}-\underline {\zeta }\,_{32}=0
\end {align}
Consider now the action of the subgroup of $\mathcal {P}$ consisting of elements $(h,a)$, where $h\in L_k$. Using eqs. (\ref{al10}), (\ref{al11}), (\ref{al19}) and (\ref{al20}) we easily conclude that to any orbit under consideration there belongs the \textquotedblleft canonical\textquotedblright point:
\begin {align} 
\label {al21}
\underline {\zeta }\,_\mu &=k_\mu \nonumber\\
\underline {\zeta }\,_{\mu \nu }&=\begin {cases} 0,&(\mu \nu )\neq (12),(21)\\-s, &(\mu \nu )=(12)\\s, & (\mu \nu )=(21)\end {cases}
\end {align}
Note that at this point 
\begin {align} 
\label {al22}
\underline {w }\,^\mu =sk^\mu 
\end {align}
Both sides of eq. (\ref{al22}) are fourvectors under the coadjoint action of Poincare group; therefore, anticipating slightly the notation,
\begin {align} 
\label {al23}
w ^\mu =sp^\mu 
\end {align} 
We see that the orbit is uniquely characterized by the single parameter $s$ which describes helicity. \\
\\
Let $\mathcal {P}_k\subset \mathcal {P}$ be the stability subgroup of canonical point (\ref{al21}) under the coadjoint action of Poincare group. It consists of the elements $\big(h,a(h)\big)$ where $h=DR\in L_k$ and $a(h)$ is defined by
\begin {align} 
\label {al24}
a^0=a^3\,-\,\text{arbitrary},\quad a^1=\frac{s}{\sqrt{2}k}d_2, \quad a^2=\frac {-s}{\sqrt{2}k}d_1
\end {align} 
with $d_{1,2}$ being defined by eq. (\ref{al4}) \\
\\
The Lie algebra of $\mathcal {P}_k$ is spanned by
\begin {align} 
\label {al25}
E_1\equiv M_{01}+M_{31}+\frac{s}{\sqrt {2}k}P_2
\end {align}
\begin {align} 
\label {al26}
E_2\equiv M_{02}-M_{23}-\frac{s}{\sqrt {2}k}P_1
\end {align}
\begin {align} 
\label {al27}
P_+=\frac{1}{\sqrt{2}}(P_0+P_3)
\end {align}
\begin {align} 
\label {al28}
J=M_{12}
\end {align}
The relevant commutation rules read
\begin {align} 
\label {al29}
[J,\,E_i]=i\epsilon _{ik}E_k
\end {align}
\begin {align} 
\label {al30}
[E_i,\,E_k]=\frac{2i\cdot s}{k}\epsilon _{ik}P_+
\end {align}
\begin {align} 
\label {al31}
[P_+,\,\cdot ]=0
\end {align}
$\mathcal {P}_k$ is, therefore, the centrally extended $E(2)$ group.\footnote{This interesting property can be also inferred from the discussions presented in Refs. \cite{b14}, \cite{b22}, \cite{b23} and, in form of symmetry of Wess-Zumino-like action, from Ref. \cite{b59}} $L_k$ is its image under the canonical homomorphism $\mathcal {P}\rightarrow SO(3,1)$. The orbit under consideration is isomorphic to the coset manifold:
\begin {align} 
\label {al32}
V=\mathcal {P}/\mathcal{P}_k
\end {align}
In order to parametrize $V$ let us consider an arbitrary element $(\Lambda,a)\in \mathcal {P}$. First, we decompose $\Lambda$ according to eq. (\ref{al2})
\begin {align} 
\label {al33}
\Lambda=B\cdot (DR)\equiv B\cdot h
\end {align}
Let $y^\mu =(0,\vec y)$; consider the decomposition
\begin {align} 
\label {al34}
(\Lambda,a)=(B,y)\cdot \bigl(h,a(h)\bigr)
\end {align}
where $a(h)$ is given by eq. (\ref{al24}). Eqs. (\ref{al33}) and (\ref{al34}) imply
\begin {align} 
\label {al35}
a^\mu =\bigl(Ba(h)\bigr)^\mu +y^\mu 
\end {align}
Eqs. (\ref{al33}) and (\ref{al35}) can be solved to yield $h$, $a(h)$, $B$ and $y$ (the solution is unique but somewhere singular due to the nontriviality of the relevant bundle, as mentioned above). The pair $(B,y)$ parametrizes the coset manifold $V$ and, consequently, the coadjoint orbit
\begin {align} 
\label {al36}
(\zeta _\mu ,\zeta _{\mu \nu })=Ad^*_{(B,y)}(\underline {\zeta }\,_\mu ,\underline {\zeta}\,_{\mu \nu })
\end {align}
with $\underline {\zeta }\,_\mu $, $\underline {\zeta}\,_{\mu \nu }$ given by eq. (\ref{al21}). Finally, we parametrize $B$ as follows:
\begin {align} 
\label {al37}
p^\mu =\Lambda^\mu _{\phantom{\mu }+}k^+=\Lambda^\mu _{\phantom{\mu }+}\cdot \sqrt {2}\cdot k, \quad p^\mu p_\mu =0
\end {align}
Using eqs. (\ref{al36}) and (\ref{al37}) one easily finds
\begin {align} 
\label {al38}
\zeta_\mu =p_\mu 
\end {align}
\begin {align} 
\label {al39}
\zeta_{12} =y_1p_2-y_2p_1
\end {align}
\begin {align} 
\label {al40}
\zeta_{23}=y_2p_3-y_3p_2+\frac{sp_1}{\sqrt {2}p^+}
\end {align}
\begin {align} 
\label {al41}
\zeta_{31}=y_3p_1-y_1p_3+\frac{sp_2}{\sqrt {2}p^+}
\end {align}
\begin {align} 
\label {al42}
\zeta_{01}=-y_1p_0+\frac{sp_2}{\sqrt {2}p^+}
\end {align}
\begin {align} 
\label {al43}
\zeta_{02}=-y_2p_0-\frac{sp_1}{\sqrt {2}p^+}
\end {align}
\begin {align} 
\label {al44}
\zeta_{03}=-y_3p_0
\end {align}
We see that the classical massless particles define the nonlinear realization of the Poincare group corresponding to the stability subgroup $\mathcal {P}_k$. The variables $\vec y$ and $\vec p$ provide independent coordinates on coadjoint orbit/coset manifold (they are Goldstone or preferred variables in terminology of Ref. \cite{b50}). Their transformation properties are derived either from coadjoint action of Poincare group on $\zeta _\mu $ and $\zeta _{\mu \nu }$ or from its left action on the coset manifold $\mathcal {P}/\mathcal {P}_k$. The relevant transformation rules can be described as follows. The momentum variables $p_\mu $ transform separately. Translations act trivially on them while the action of infinitesimal Lorentz transformations $\Lambda^\mu _{\phantom {\mu }\nu }=\delta ^\mu _{\phantom {\mu }\nu }+\omega ^\mu _{\phantom {\mu }\nu }$ reads
\begin {align} 
\label {al45}
\delta p_k=\beta _k\mid \vec p\mid +\epsilon _{klm}\omega _lp_m
\end {align}
where $\beta _k\equiv \omega _{k0}, \; \epsilon _{ikl}\omega _l=\omega _{ik}$. The translation subgroup is the kernel of the realization on momentum variables which is the nonlinear realization of Lorentz group determined by the $E(2)$ subgroup (stability subgroup of $k^\mu )$. In spite of the fact that the third axis plays a distinguished role the realization linearizes on rotations. The $y$-variables transform in a more complicated way. Consider again Lorentz transformations. They read
\begin {align} 
\label {al46}
\delta \vec y=\vec \omega \times \vec y-(\vec \beta \cdot \vec y)\frac{\vec p}{\mid \vec p\mid} +\vec \bigtriangledown _p\Psi (p)
\end {align}
where
\begin {align} 
\label {al47}
\Psi  (p)=s\Biggl(\frac {\omega _1p_1+\omega _2p_2+\beta _1p_2-\beta _2p_1}{\sqrt {2}p^+}\Biggr)
\end {align}
It is not difficult to check that the transformations (\ref{al45})-(\ref{al47}) are canonical
\begin {align} 
\label {al48}
\delta (\cdot )=\Bigl\{(\cdot ),\,\frac{1}{2}\omega ^{\mu \nu }\zeta _{\mu \nu }\Bigr\}
\end {align}
As a next step we find the Poisson brackets for $y's$ and $p's$. Using eqs. (\ref{al12})-(\ref{al14}) and (\ref{al38})-(\ref{al44}) one computes
\begin {align} 
\label {al49}
\{y_i,\,y_k\}=0
\end {align}
\begin {align} 
\label {al50}
\{y_i,\,p_k\}=\delta _{ik}
\end {align}
\begin {align} 
\label {al51}
\{p_i,\,p_k\}=0
\end {align}
In terms of $y's$ rotations linearize only on the subgroup of rotations around the third axis. However, one can make things explicitly rotationally invariant by passing to the coordinates $\vec x$ defined as follows:
\begin {align} 
\label {al52}
y_1=x_1+\frac{s p_2}{\sqrt {2} p^+p^0}
\end {align}
\begin {align} 
\label {al53}
y_2=x_2-\frac{s p_1}{\sqrt {2} p^+p^0}
\end {align}
\begin {align} 
\label {al54}
y_3=x_3
\end {align}
Then we find
\begin {align} 
\label {al55}
\zeta _{0i}=-p_0x_i
\end {align}
\begin {align} 
\label {al56}
\zeta _{ij}=x_ip_j-x_jp_i+\frac{s \epsilon _{ijk}p_k}{p^0}
\end {align}
The price one has to pay for simplifying the transformation properties is that the new variables are no longer Darboux ones. In fact, new Poisson brackets read
\begin {align} 
\label {al57}
\{x_i,\,x_j\}=-\frac{s \epsilon _{ijk}p_k}{(p^0)^3}
\end {align}
\begin {align} 
\label {al58}
\{x_i,\,p_j\}=\delta _{ij}
\end {align}
\begin {align} 
\label {al59}
\{p_i,\,p_j\}=0
\end {align}
In terms of new variables rotations, as it has been mentioned above, take the standard form. On the other hand, the boosts read
\begin {align} 
\label {al60}
\delta x_i=-(\beta _j\cdot x_j)\frac{p_i}{\mid \vec p\mid }+\frac{s \epsilon _{ijk}\beta _j p _k}{\mid \vec p\mid ^2}
\end {align}
\begin {align} 
\label {al61}
\delta p_i=\beta _i\mid \vec p\mid 
\end {align}
For completeness let us write out the action of translation subgroup $(I,a)$. It reads
\begin {align} 
\label {al62}
\delta x_i&=a_i-a_0\frac{p_i}{\mid \vec p\mid}\nonumber\\
\delta p_i&=0
\end {align}
This agrees with the identification $H=p^0=\mid \vec p \mid $.
\section {Poincare symmetry}
\par To derive the symmetry transformations we note that our symmetry is a dynamical one: the Hamiltonian belongs to the Lie algebra of symmetry group and, in general, does not Poisson-commute with other generators. Their time evolution is given by the one parameter subgroup of adjoint transformations generated by the Hamiltonian. Expressing the initial $(t=0)$ generators in terms of actual ones yields the conserved charges which generate the symmetry. In our case the new conserved generators read
\begin {align} 
\label {al63}
\tilde \zeta _\mu =\zeta _\mu 
\end {align}
\begin {align} 
\label {al64}
\tilde \zeta _{ij} =\zeta _{ij}
\end {align}
\begin {align} 
\label {al65}
\tilde \zeta _{0i} =\zeta _{0i}+\zeta _it
\end {align}
\par By virtue of eqs. (\ref{al63})-(\ref{al65}) we conclude that the symmetry transformations corresponding to the boosts are modified according to
\begin {align} 
\label {al66}
\delta x_i=-(\beta _kx_k)\frac{p_i}{\mid \vec p \mid }+\beta _it+\frac {s \epsilon _{ijk}\beta _jp_k}{\mid \vec p\mid ^2}
\end {align}
\par The same result is obtained by applying the original transformations (\ref{al60}) and (\ref{al61}) to initial variables and propagating them to the moment $t$ with the help of equations of motion.
\par Concluding, the symmetry transformations are obtained through nonlinear action of Poincare group on the coset manifold defined by the subgroup of Poincare group related to $\mathcal {P}_k$ by a time-dependent internal automorphism generated by the Hamiltonian. In other words, let $\vec x=\vec x(t,\vec x_0,\vec p_0)$, $\vec p=\vec p(t,\vec x_0,\vec p_0)$ be the solution to the equations of motion; the change of variables $(\vec x,\vec p,t)\rightarrow (\vec x_0,\vec p_0,t)$ yields the nonlinear realization with $\vec x_0$, $\vec p_0$ being the preferred variables parametrizing $\mathcal {P}/\mathcal {P}_k$ while $t$ is the adjoint variable \cite{b50} transforming trivially under the action of $\mathcal {P}_k$.
\par Transformation rules (\ref{al66}) can be put in yet another form. Within the Hamiltonian formalism the symmetry transformations do not involve the redefinition of time. The symmetries including the change of time variable are accommodated by recomputing the values of dynamical variables back to initial time with the help of canonical equations of motion; for any dynamical variable $\eta $ the relation between the Hamiltonian and Lagrangian form of symmetries reads $\delta _H\eta =\delta _L\eta -\dot \eta \delta t$. Keeping this in mind we rewrite the transformation rules (\ref{al66}) as
\begin {align} 
\label {al67}
\delta t=\beta _k x_k
\end {align}
\begin {align} 
\label {al68}
\delta x_i=\beta _it+\frac{s \epsilon _{ijk}\beta _jp_k}{\mid \vec p\mid^2 }
\end {align}
\begin {align} 
\label {al69}
\delta p_i=\beta _i\mid\vec p \mid 
\end {align}
For $s =0$ one arrives at the standard Lorentz transformation rules. However, for $s \neq 0$ the above transformation rules close only \textquotedblleft on-shell\textquotedblright \cite{b16}. Indeed, the relevant differential boost generators read
\begin {align} 
\label {al70}
M_{0k}=i\Biggl(x_k\frac{\partial }{\partial t}+t\frac{\partial }{\partial _xk}+\frac{s \epsilon _{klj}p_l}{\mid \vec p\mid ^2}\,\frac{\partial }{\partial x_j}+\mid \vec p \mid \frac {\partial }{\partial p_k}\Biggr)
\end {align}
The corresponding commutation rule takes the form
\begin {align} 
\label {al71}
[M_{0k},\,M_{0m}]=-iM_{km}-\frac{2s \epsilon _{kml}p_l}{\mid \vec p\mid^2 } \Biggl(\frac{\partial }{\partial t}+\frac{p_j}{\mid \vec p\mid }\,\frac{\partial }{\partial x_j}\Biggr )
\end {align}
and reduces to the standard form on trajectories $x_k-\frac{p_kt}{\mid \vec p \mid }=const$.
\section {Quantum theory}
\par It is easy to quantize the classical theory formulated above. We start with diagonalizing the momenta yielding the momentum representation. As the momentum variables transform in a standard way it is convenient to use the explicitly invariant scalar product
\begin {align} 
\label {al72}
(f,g)=\int\frac{d^3\vec p}{2\mid \vec p\mid } \overline {f(\vec p)} g(\vec p)
\end {align}
\par Due to the canonical relations (\ref{al49})-(\ref{al51}) $y's$ are basically $p$-derivatives. However, we should take into account the hermicity condition with respect to the scalar product (\ref{al72}). Therefore,
\begin {align} 
\label {al73}
\vec y=\sqrt {\mid \vec p \mid }\Biggl(\frac{i\partial }{\partial \vec p}\Biggr)\frac{1}{\sqrt{\mid \vec p \mid }}=\frac{i\partial }{\partial \vec p}-\frac{i\vec p}{2\mid \vec p\mid^2 }
\end {align}
\par Now, one can construct generators according to the equations (\ref{al38})-(\ref{al44}); to this end one has to perform symmetrization $y_ip_0\rightarrow \frac{1}{2}(y_ip_0+p_0y_i)$. The resulting generators read 
\begin {align} 
\label {al74}
P_\mu =p_\mu 
\end {align}
\begin {align} 
\label {al75}
M_{12}=i\Biggl(p_2\frac{\partial }{\partial p_1}-p_1\frac{\partial }{\partial p_2}\Biggr)-s
\end {align}
\begin {align} 
\label {al76}
M_{23}=i\Biggl(p_3\frac{\partial }{\partial p_2}-p_2\frac{\partial }{\partial p_3}\Biggr)+\frac {sp_1}{p_0-p_3}
\end {align}
\begin {align} 
\label {al77}
M_{31}=i\Biggl(p_1\frac{\partial }{\partial p_3}-p_3\frac{\partial }{\partial p_1}\Biggr)+\frac {sp_2}{p_0-p_3}
\end {align}
\begin {align} 
\label {al78}
M_{01}=-i\mid\vec p \mid \frac{\partial }{\partial p_1}+\frac{sp_2}{p_0-p_3}
\end {align}
\begin {align} 
\label {al79}
M_{02}=-i\mid\vec p \mid \frac{\partial }{\partial p_2}-\frac{sp_1}{p_0-p_3}
\end {align}
\begin {align} 
\label {al80}
M_{03}=-i\mid\vec p \mid \frac{\partial }{\partial p_3}
\end {align}
\par It is easy to check that $M_{\mu \nu }$ and $P_\mu $ obey Poincare algebra. If we demand that it integrates to the representation of universal covering $ISL(2,C)$ of Poincare group $s$ must be integer or halfinteger. The above representation coincides with that given, for example, in Refs. \cite{b51} or \cite{b52}.
\par We see that the straightforward quantization of the Hamiltonian system built on the coadjoint orbits characterized by $\zeta _\mu \zeta ^\mu =0$, $w _\mu w ^\mu =0$ yields irreducible representations corresponding to massless particles of arbitrary helicity $s$.
\section {Action principle}
\par It is well-known that the Kirillov form defining Poisson brackets on coadjoint orbit is related to the Cartan forms on the relevant coset manifold \cite{b53}. Applying the prescription given in \cite{b53} to the case of Poincare symmetry we define
\begin {align} 
\label {al81}
\Omega (p,y)\equiv \Omega ^\mu (p,y)\underline {\zeta }\,_\mu +\Omega ^{\mu \nu }(p,y)\underline {\zeta }\,_{\mu \nu }
\end {align}
where $(p,y)$ parametrize the coset manifold and 
\begin {align} 
\label {al82}
(p,y)^{-1} d(p,y)=i\Omega ^\mu (p,y)P_\mu +i\Omega ^{\mu \nu }(p,y)M_{\mu \nu }
\end {align}
Then
\begin {align} 
\label {al83}
\tilde \Omega \equiv d\Omega 
\end {align}
is the relevant Kirillov form. Explicit computation yields
\begin {align} 
\label {al84}
\Omega = -p^idy^i=-p^idx^i+\alpha ^i(\vec p)dp^i
\end {align}
with
\begin {align} 
\label {al85}
\vec \alpha (\vec p)=s \Biggl(\frac{-p^2}{p^0p^+},\,\frac{p^1}{p^0 p^+}, \, 0\Biggr)
\end {align}
being the vector potential of the monopole. \\
\par According to the general theory the action functional yielding correct equations of motion reads
\begin {align} 
\label {al86}
S=\int \bigl(-\Omega -Hdt\bigr)
\end {align}
leading to
\begin {align} 
\label {al87}
S=\int \bigl(\vec p\cdot \dot {\vec y}-\mid \vec p\mid \bigr)dt=\int\bigl(\vec p\cdot \dot {\vec x}-\mid \vec p\mid -\vec \alpha (\vec p)\cdot \dot {\vec p}\bigr)dt
\end {align}
\par The first form of the action integral confirms the conclusion that $(\vec y,\vec p)$ are Darboux variables. Let us note that it does not depend on the helicity $s$. Before entering more sophisticated aspects of action principle let us make some remarks. The textbook action for massive relativistic particle reads
\begin {align} 
\label {al88}
S=-m\int ds=-m\int\sqrt{1-\dot {\vec y}\,^2}dt
\end {align}
\par The $m\rightarrow 0$ limit cannot be taken directly. However, one can pass to the Hamiltonian form which is straightforward for $m\neq 0$ and yields
\begin {align} 
\label {al89}
S=-\int p_\mu dy^\mu , \quad y^0=t, \quad p_\mu p^\mu =0
\end {align}
\par It is now easy to take the limit $m\rightarrow 0$ which gives eq.  (\ref{al87}). In terms of $y$, $p$ variables the action has the universal form as it does not depend on the helicity value. The latter enters only the transformation rule. Let us write it in \textquotedblleft Lagrangian\textquotedblright form:
\begin {align} 
\label {al90}
\delta y^0=\vec \beta \cdot \vec y
\end {align}
\begin {align} 
\label {al91}
\delta \vec y=\vec \omega \times \vec y+\vec \beta y^0+\vec \bigtriangledown _p\Psi  (p)
\end {align}
\begin {align} 
\label {al92}
\delta \vec p=\vec \omega \times \vec p+\vec \beta p^0
\end {align}
\begin {align} 
\label {al93}
\delta p^0=\vec \beta \cdot \vec p
\end {align}
$\Psi  (p)$ is proportional to the helicity and this is the only term where $s$ enters. The additional contribution to the action integrand reads
\begin {align} 
\label {al94}
\vec p d\bigl(\vec \bigtriangledown _p\Psi (p)\bigr)&=d\bigl(\vec p\cdot \vec \bigtriangledown _p\Psi (p)\bigr)-d \vec p\cdot \vec \bigtriangledown _p\Psi (p)=\nonumber\\
&=d \bigl(\vec p\cdot \vec \bigtriangledown _p\Psi (p)-\Psi (\vec p)\bigr)
\end {align}
which proves the invariance of action principle.
\par Let us consider in more detail the action principle and space-time description of massless particles. It is easy to note that in the case of dynamics on coadjoint orbit all dynamical variables can be expressed in terms of the generators of canonical transformations representing the symmetries; the quantum counterpart of this statement is that in the case of irreducible representation all observables are (at least in principle) expressible in terms of symmetry generators. In particular, the particle coordinates can be (in more or less sensible way) written in terms of generators (cf. eqs. (\ref{al38})-(\ref{al44})). We saw that $\vec y\,'s$ and $\vec x\,'s$, defined in this way, have slightly unusual transformation properties. However, as we discussed in Introduction, these transformation rules can be, to some extent, justified by considering various physical phenomena. On the other hand, the space-time coordinates transforming in the standard way enter almost inevitably when interactions between relativistic particles are considered. As it is strongly advocated by Weinberg \cite{b54}, the Lorentz-covariant interacting theory can be (at least, most easily) constructed by starting with covariant and causal fields on standard space-time. Moreover, in certain limit one obtains a quantum particle propagating in classical field (mostly electromagnetic) depending on space-time variables transforming according to the usual rule. One can pose the question whether and how the space-time variables transforming according to the formula $x\rightarrow \Lambda x+a$ can be built into the theory. New formalism should be equivalent to the one based on coadjoint orbits. Therefore, all dynamical variables should be constructed out of group elements. Assume the global symmetry is identified with (say) left action of the group on itself. From the form of nonlinear group action we conclude that our dynamics must be invariant under the right action of stability subgroup viewed as the gauge group. In fact, the relevant dynamical variables parametrize the coset space. If one works with the variables parametrizing the whole group, those corresponding to the subgroup must be redundant and should be eliminated by a symmetry transformations. Writing schematically the nonlinear action on coset space
\begin {align} 
\label {al95}
gw=w'h(w,g)
\end {align}
we see that in order to eliminate the subgroup variables one has to act from the right with the stability subgroup elements which are generally time-dependent; we are dealing with gauge symmetry.
\par It is quite easy to write out the action integral on group manifold which is invariant under the global action of this group by left multiplication and the local action of some its subgroup by right multiplication provided this subgroup is the stability group of some point on coadjoint action. Let $G$ be a Lie group, $H\subset G$ its subgroup leaving invariant the element $\underline {\xi }\,_\alpha $ of dual space to Lie algebra of $G$. Writing the Cartan-Maurer form as
\begin {align} 
\label {al96}
g^{-1}dg=i\eta  ^\alpha (g)A_\alpha 
\end {align}
where $A_\alpha $ are the generators of $G$ and putting 
\begin {align} 
\label {al97}
\eta  (g)\equiv \eta  ^\alpha (g)\underline{\xi }\,_\alpha 
\end {align}
one easily finds that $\omega (g)$ is invariant under the global left action of $G$ and, up to a total differential, under the local right action of $H$. Therefore, the first-order action
\begin {align} 
\label {al98}
S=\int\bigl(-\eta  (g)\bigr)
\end {align}
defines invariant dynamics of $G/H$ (because of the gauge symmetry under the right action of $H$).
\par Let apply the above construction to the Poincare group. One has
\begin {align} 
\label {al99}
(\Lambda,z)^{-1}d(\Lambda,z)=(\Lambda^{-1}d\Lambda,\Lambda^{-1}dz)
\end {align}
\par Keeping in mind the form of \textquotedblleft canonical\textquotedblright point (\ref{al21}) we arrive easily at the following form of invariant action
\begin {align} 
\label {al100}
S=-\int\Bigl(k\bigl(\Lambda_\mu^{\phantom {\mu }0}-\Lambda_\mu ^{\phantom {\mu }3}\bigr)dz^\mu -\frac {is}{2} Tr\bigl(J\Lambda^{-1}d\Lambda\bigr)\Bigr)
\end {align}
with $J=M_{12}$. This Wess-Zumino like action was considered in Refs. \cite{b55}-\cite{b59}. (see also \cite{b60}). It posses the expected symmetries under:
\begin {itemize}
\item [--] the global Poincare transformations:
\begin {align} 
\label {al101}
(\tilde \Lambda,a):\;(\Lambda,z)\longrightarrow (\tilde \Lambda\Lambda,\tilde \Lambda z+a)
\end {align}
\item [--] the local $\mathcal{P}_k$ transformations:
\begin {align} 
\label {al102}
\delta \Lambda =i \theta ^a(t)\Lambda\tilde E_a+i\varphi (t)\Lambda M
\end {align}
\begin {align} 
\label {al103}
\delta z^\mu=\frac{s}{k}\Bigl(\theta ^1(t)\Lambda^\mu _{\phantom{\mu} 2}-\theta ^{(2)}(t)\Lambda^\mu _{\phantom{\mu }1}\Bigr)+a(t)(\Lambda^\mu _{\phantom {\mu }0}+\Lambda^\mu _{\phantom {\mu }3})
\end {align}
\begin {align} 
\label {al104}
\tilde E_1\equiv M_{01}+M_{31}, \quad \tilde E_2\equiv M_{02}-M_{23}
\end {align}
\end {itemize}
\par The first order action (\ref{al100}) exhibits gauge symmetry. A careful analysis of the emerging constraints leads, via Dirac method, to the conclusion that it describes, as expected, the massless helicity $s$ particles (cf., for example, Ref. \cite{b59} ). One can also proceed by fixing an appropriate gauge. To this end we recall the decomposition (\ref{al2})-(\ref{al8}) together with the identification (\ref{al37}). It follows that one can fix the gauge such that $\Lambda=B$ with matrix elements being parametrized by fourmomentum $p^\mu $. Moreover, the parameter function $a(t)$ can be chosen in such a way that $z^0\equiv t$ (in other words the invariant evolution parameter can be replaced by time).
\par Fixing the gauge as above we arrive at the following simple action
\begin {align} 
\label {al105}
S=\int\bigl(\vec p\cdot \dot  {\vec y}-\mid \vec p\mid \bigr)dt
\end {align}
which coincides with eq. (\ref{al89}); here $\vec y$ denotes the spatial part of gauge-transformed $z^\mu $.
\par The findings of previous sections can be now rephrased as follows. The gauge fixing condition breaks the explicit global Poincare invariance. The Poincare transformation must be supplemented by an appropriate gauge transformation which restores the gauge. This makes the final transformation rule more complicated. 
\par The action (\ref{al100}) yields the following equations of motion \cite{b55}-\cite{b59}
\begin {align} 
\label {al106}
\frac {dp^\mu }{d\tau }=0, \quad p^\mu \equiv \Lambda^\mu _{\phantom {\mu }\nu }k^\nu , \quad k^\nu =(k,0,0,k)
\end {align}
\begin {align} 
\label {al107}
\frac {dz^\mu }{d\tau }p^\nu -\frac{dz^\nu }{d\tau }p^\mu -\frac{s}{2}\frac{d}{d\tau }\Bigl(Tr(\Lambda J \Lambda^{-1}M^{\mu \nu })\Bigr)=0
\end {align}
\par Eqs. (\ref{al106}) and (\ref{al107}) are invariant under the action of gauge transformations (\ref{al102})-(\ref{al104}). As a result one can choose $\theta ^a(t)$ and $\varphi (t)$ in such a way that $\Lambda=B$. The last term on the right hand side of eq. (\ref{al107}) vanishes and we find
\begin {align} 
\label {al108}
\frac {dz^\mu }{d\tau }p^\nu -\frac{dz^\nu }{d\tau }p^\mu =0
\end {align}
which implies 
\begin {align} 
\label {al109}
p^\mu =\sigma (\tau )\frac{dz^\mu }{d\tau }
\end {align}
The coefficient $\sigma (\tau )$ can be changed by the residual freedom in eq. (\ref{al103}).
\begin {align} 
\label {al110}
\delta z^\mu =\frac{a(\tau )}{k}p^\mu 
\end {align}
In particular, one can choose $\sigma (\tau )=p^0$; then $\tau =z^0$ and $\frac{dz^i}{dz^0}=\frac{p^i}{p^0}$ yielding standard equations of motion.
\par Let us now consider the coupling to the external electromagnetic field. The minimal coupling is achieved by adding the term $eA_\mu (z)\dot z^\mu $ to the Lagrangian. The $z$ variables transform standardly under Poincare group so $A_\mu (z)$ have the standard meaning. The action takes the form
\begin {align} 
\label {al111}
S=\int\Bigl(-(\Lambda_{\mu 0}+\Lambda_{\mu 3})kdz^\mu +eA_\mu (z)dz^\mu -\frac{is}{2}Tr(J\Lambda^{-1}d\Lambda)\Bigr)
\end {align}
\par Note that the above action, while preserving the standard gauge invariance related to electromagnetic coupling seems to break the gauge symmetry related to the right action of stability subgroup. This could imply that some of the gauge degrees of freedom become real dynamical variables. The related ambiguity in including the interaction has been discussed in the Refs. \cite{b22}, \cite{b23}. The problem of interaction will be treated in more detail in the forthcoming paper \cite{b61}.
\section {Concluding remarks}
\par We have discussed the dynamics of classical massless particles within the well-known method of coadjoint orbits. As expected, we arrived at the Hamiltonian description of massless particles which after canonical quantization leads to the unitary representations of Poincare group corresponding to zero mass and arbitrary helicity. By interpreting the coadjoint action in terms of nonlinear realization of Poincare group we provided a natural explanation of \textquotedblleft exotic\textquotedblright transformation rules for particle \textquotedblleft coordinates\textquotedblright. These rules, whatever exotic they are, agree with what is expected on the basis of simple relativistic considerations \cite{b15}, \cite{b16}, \cite{b35}-\cite{b37}.
\par By introducing the additional gauge degrees of freedom (suggested by the general form of nonlinear realizations) one can construct space-time coordinates transforming according to the standard rules. The existence of such coordinates is crucial as far as interaction is concerned. It is well known (see Ref. \cite{b54} for beautiful explanation) that, in order to define covariant, unitary and causal interaction one has to introduce local fields depending on space-time coordinates transforming according to the standard representation of Poincare group. These fields provide the building blocks of fully consistent quantum interacting theory.
\par In certain limit one obtains the dynamics of quantum particle in external classical field \cite{b54}. It can be described by an appropriate wave equation. The most interesting case is that of the motion of quantum particle in classical electromagnetic field. The latter has a natural description in terms of fields depending on space-time coordinates transforming in the standard way under the action of Poincare group. The covariant wave equation is not only described by the very choice of the unitary representation of Poincare group but also by the choice of the representation of Lorentz group acting on covariant field carrying this representation. It would be desirable to find the appropriate description of the interaction with electromagnetic field on the classical level within the scheme described in the present paper. 
\par Another interesting question concerns the relation between the present formalism for massless particles with nonzero helicities and the massive ones with spin based on twistor theory \cite{b62}-\cite{b64}. Also the relation with other models of massless particles \cite{b65}-\cite{b68} is worth of study.
\\
\\
{\bf Acknowledgments}
Piotr Kosi\'nski gratefully acknowledges fruitful discussion and kind correspondence with P. Horvathy and J. Lukierski. The research was supported by the grant of National Science Center number DEC-2013/09/B/ST2/02205.
\\
\\

\end{document}